\documentclass[amsmath,amssymb,amsfonts,aps,prl,twocolumn,superscriptaddress]{revtex4-1}

\usepackage{color}
\usepackage{graphicx}
\usepackage{dcolumn}
\usepackage{bm}

\renewcommand{\Re}{{\rm Re}}

\newcommand{\GammaK}{\Gamma_{\rm K}}
\newcommand{\DeltaK}{\Delta_{\rm K}}

\newcommand{\TK}{T_{\rm K}}
\newcommand{\TKN}{T_{\rm K,N}}
\newcommand{\kB}{k_{\rm B}}

\newcommand{\emp}{{\rm emp}}

\newcommand{\didv}{{\rm d}I/{\rm d}V}

\usepackage[normalem]{ulem}

\begin{document}

\title{Demonstrating Kondo behavior by temperature-dependent scanning tunneling spectroscopy}

\author{Elia Turco}
\email{elia.turco@empa.ch}
\affiliation{nanotech@surfaces Laboratory, Empa - Swiss Federal Laboratories for Materials Science and Technology, 8600 Dübendorf, Switzerland}

\author{Markus Aapro}
\email{markus.aapro@aalto.fi}
\affiliation{Department of Applied Physics, Aalto University, 00076, Aalto, Finland}

\author{Somesh C. Ganguli}
\affiliation{Department of Applied Physics, Aalto University, 00076, Aalto, Finland}

\author{Nils Krane}
\affiliation{nanotech@surfaces Laboratory, Empa - Swiss Federal Laboratories for Materials Science and Technology, 8600 Dübendorf, Switzerland}

\author{Robert Drost}
\affiliation{Department of Applied Physics, Aalto University, 00076, Aalto, Finland}

\author{Nahual Sobrino}
\affiliation{Departamento de Polímeros y Materiales Avanzados: Física, Química y Tecnología,
Universidad del País Vasco UPV/EHU, Av. Tolosa 72, E-20018 San Sebastián, Spain}

\author{Annika Bernhardt}
\author{Michal Juríček}
\affiliation{Department of Chemistry, University of Zurich, Winterthurerstrasse 190, 8057 Zurich, Switzerland}

\author{Roman Fasel}
\affiliation{nanotech@surfaces Laboratory, Empa - Swiss Federal Laboratories for Materials Science and Technology, 8600 Dübendorf, Switzerland}
\affiliation{Department of Chemistry, Biochemistry and Pharmaceutical Sciences, University of Bern, 3012 Bern, Switzerland}

\author{Pascal Ruffieux}
\affiliation{nanotech@surfaces Laboratory, Empa - Swiss Federal Laboratories for Materials Science and Technology, 8600 Dübendorf, Switzerland}
\email{pascal.ruffieux@empa.ch}

\author{Peter Liljeroth}
\affiliation{Department of Applied Physics, Aalto University, 00076, Aalto, Finland}
\email{peter.liljeroth@aalto.fi}

\author{David Jacob}
\email{david.jacob@ehu.es}
\affiliation{Departamento de Polímeros y Materiales Avanzados: Física, Química y Tecnología,
Universidad del País Vasco UPV/EHU, Av. Tolosa 72, E-20018 San Sebastián, Spain}
\affiliation{IKERBASQUE, Basque Foundation for Science, Plaza Euskadi 5, E-48009 Bilbao, Spain}

\begin{abstract}
  The Kondo effect describes the scattering of conduction electrons by magnetic
  impurities, manifesting as an electronic resonance at the Fermi energy with a
  distinctive temperature evolution. In this letter, we present a critical
  evaluation of the current methodology employed to demonstrate Kondo behavior
  in transport measurements, underscoring the limitations of established
  theoretical frameworks and the influence of extrinsic broadening. We
  introduce a novel approach for analyzing spectroscopic indicators of the
  Kondo effect, employing the Hurwitz-Fano lineshape as a model for the Kondo
  resonance in the presence of extrinsic broadening. Through precise scanning
  tunneling spectroscopy measurements on an exemplary spin-1/2 Kondo system,
  phenalenyl on Au(111), we demonstrate the efficacy of our proposed protocol
  in extracting accurate intrinsic Kondo linewidths from finite-temperature
  measurements. The extracted linewidths exhibit a robust fit with a recently
  derived expression for the temperature-dependent intrinsic Kondo linewidth,
  providing compelling evidence for the validity of the underlying theory.
\end{abstract}

\maketitle


The Kondo effect is one of the most enigmatic phenomena in condensed matter physics and one of the hallmarks of strong
electronic correlations~\cite{Kondo:PTP:1964,Hewson:book:1997,Fulde:book:1995,Fazekas:book:1999}. It occurs when a
magnetic impurity interacts with the conduction electrons in a metallic host. Below a characteristic temperature,
called the Kondo temperature, $\TK$, the impurity spin is effectively screened by the formation of a total spin-singlet
state with the surrounding conduction electrons. As the low-temperature dynamics of individual Kondo impurities are
governed by the energy scale $\kB\TK$, its magnitude has a major influence on the ground states and quantum phase
diagrams of strongly correlated materials, such as Kondo lattices and heavy fermion
systems~\cite{Coleman:book_chapter:2007,Loehneysen:RMP:2007,Keimer:NPhys:2017,Zhao:Nature:2023,Roch2008}.

The Kondo effect is signaled by the appearance of a sharp resonance at the Fermi level in the density of states of the
magnetic impurity. This can be exploited for the detection of magnetic moments in atoms and molecules adsorbed on
metallic substrates by scanning tunneling spectroscopy (STS), where it shows up as a zero-bias anomaly in the conductance
($\didv$) spectra~\cite{Madhavan:Science:1998,Li:PRL:1998,Manoharan:Nature:2000,Ujsaghy:PRL:2000,Schiller:PRB:2000}.
However, a clear-cut proof of Kondo behavior requires to discriminate the Kondo resonance from other zero-bias anomalies.
This can be achieved, for example, by measuring the temperature evolution of the resonance's
linewidth~\cite{Nagaoka:PRL:2002,Mishra:NNano:2019,Li:PRL:2020,Mishra:Nature:2021,Turco:JACSAu:2023},
which shows a characteristic universal behavior in the Kondo regime~\cite{Costi:PRL:2000,Osolin:PRB:2013}.
 
Attempts to derive an analytic expression for the temperature dependence of the Kondo peak from Fermi liquid
theory~\cite{Nagaoka:PRL:2002} were shown to be problematic due to the limitation to very low temperatures $T\ll\TK$
and energies $\omega\ll\GammaK$~\cite{Chen:PRB:2021,Jacob:PRB:2023}. As a consequence, empirical expressions for the
temperature-dependent Kondo linewidth are often used to fit experimental data, \textit{e.g.},
$\Gamma_\emp(T) = \sqrt{(\alpha \kB T)^2+2 (\kB \TKN)^2}$ where $\alpha$ was introduced as an additional fitting
parameter instead of $\alpha=\pi$ as originally defined in Ref.~\cite{Nagaoka:PRL:2002}. Due to the free parameter
$\alpha$ and different definitions for the Kondo temperature (\textit{e.g.} $\TKN = 2.77\, \TK$)~\cite{supplemental},
various forms for $\Gamma_\emp(T)$ can be found in the
literature~\cite{Zhang:NComm:2013,Khajetoorians:NNano:2015,Esat:PRB:2015,Mishra:NNano:2019,Li:PRL:2020,Turco:JACSAu:2023}.
It has also been noted that simple square-root expressions such as $\Gamma_\emp(T)$ cannot capture the universal
scaling behavior obtained in accurate numerical renormalization group calculations~\cite{Note3}.

Recently, an analytic equation for the temperature-dependent linewidth of the Kondo peak was derived from a novel
theoretical Ansatz for the renormalized self-energy that extends the temperature and energy range beyond the Fermi
liquid regime~\cite{Jacob:PRB:2023}:
\begin{equation}
  \label{eq:kondowidth}
  \Gamma(T) = \DeltaK \cdot\sqrt{ a + b\,\sqrt{1+\left(\tfrac{\tau}{\DeltaK}\right)^2} + c \,
    \left(\tfrac{\tau}{\DeltaK}\right)^2 }
\end{equation}
where $a\equiv1+\sqrt3\sim2.732$, $b\equiv2+\sqrt3\sim3.732$ and $c\equiv\sqrt3/2\sim0.866$ are constants,
$\tau\equiv\pi \kB T$ is the temperature parameter and $\DeltaK$ the width parameter of the $T=0$ Kondo peak,
related to the halfwidth via $\GammaK=2.542\,\DeltaK$, and to the Kondo temperature via
$\DeltaK=1.542\,\kB\TK$~\cite{Note1}. As shown in Ref.~\onlinecite{Jacob:PRB:2023}, Eq.~\ref{eq:kondowidth}
is in excellent agreement with numerical renormalization group calculations~\cite{Osolin:PRB:2013}. On the other hand,
experimental data of Kondo linewidths versus temperature measured by STS~\cite{Mishra:NNano:2019} could not be fitted
very well, most likely due to the presence of \emph{extrinsic} broadening mechanisms in the STS
data~\cite{Gruber:JPCM:2018}. Post-hoc removal of extrinsic broadening from the experimentally measured widths only 
slightly improved the agreement. Therefore a clear experimental proof of the theory of Ref.~\cite{Jacob:PRB:2023} is
still lacking.

\begin{figure}
  \includegraphics[width=\linewidth]{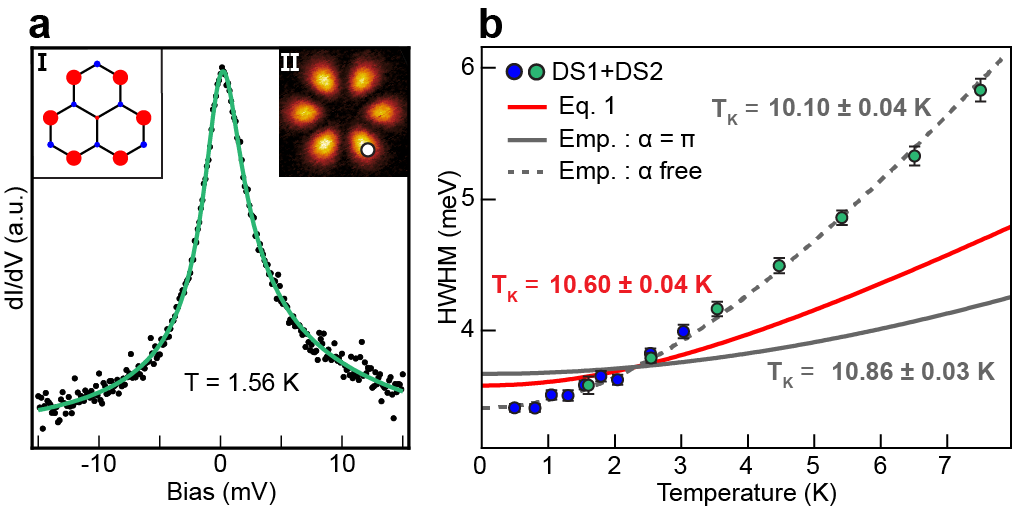}
  \caption{
    \label{fig:frotafits}
    (a) $\didv$ spectrum of phenalenyl from DS2 at {$T=1.56$K} (black dots) showing the Kondo
    resonance, fitted by Frota-Fano lineshape (green line). Insets I and II show the theoretical
    and experimental spatial distribution of the unpaired electron~\cite{Turco:JACSAu:2023}. (b)
    Kondo linewidths extracted from Frota fits to $\didv$ spectra of DS1 (blue circles) and DS2
    (green circles) versus temperature, and fits to Eq.~\ref{eq:kondowidth} (red solid line),
    and to $\Gamma_\emp(T)$ with fixed $\alpha=\pi$ (grey solid line) and with $\alpha$ as fit
    parameter (grey dashed line, $\alpha=7.45$). Error bars show the estimated standard
    deviation.
  }
\end{figure}

In this work, we demonstrate the validity of the theory reported in Ref.~\onlinecite{Jacob:PRB:2023} by accurate
temperature-dependent measurements and analysis of a prototypical spin-1/2 Kondo system, and provide an efficient
protocol to experimentally prove the Kondo nature of a zero-bias peak. To this end, we carried out low-temperature STM
experiments on phenalenyl molecules deposited on a Au(111) surface~\cite{supplemental}. The unpaired $\pi_z$ electron
of phenalenyl forms an $S=1/2$ ground state and is uniformly delocalized over six equivalent positions, as shown in the
two insets of Fig.~\ref{fig:frotafits}(a) where the spin density plot (inset I) and the experimental constant-height map
of the Kondo resonance (inset II) are reported (more details in Ref.~\onlinecite{Turco:JACSAu:2023}). The
high-resolution STS spectra shown in this work were measured with a metal STM tip and acquired as point spectra on one
of the six equivalent Kondo lobes, as depicted in inset II of Fig.~\ref{fig:frotafits}(a).  

Two distinct data sets of temperature-dependent $\didv$ spectra were acquired on two different molecules, which are
similarly adsorbed on the face centered cubic (\textit{fcc}) region of the Au(111) herringbone reconstruction. We refer
to the lower temperature data set as DS1, while DS2 is the data set from 1.56K to 7.5K. For each temperature the Kondo
peak was fitted with the Frota-Fano lineshape (example shown in Fig.~\ref{fig:frotafits}a), taking quantum interference into account via
the Fano phase $\phi$~\cite{Frota:PRB:1992,Prueser:PRL:2012,Frank:PRB:2015},
$F(V) = F_0 \cdot \Re\left[ e^{i\phi}/\sqrt{1+iV/\Delta}  \right]$ where $F_0$ is the amplitude, and $\Delta$ the Frota
width parameter related to the halfwidth of the Frota-Fano lineshape by
$\Gamma=\sqrt{3+\sqrt{12}}\cdot\Delta=2.542\cdot\Delta$. The Frota-Fano lineshape yields good fits for all
temperatures~\cite{supplemental}. The obtained halfwidths of the Frota fits are shown in Fig.~\ref{fig:frotafits}(b)
for both data sets. The two data sets overlap neatly in the temperature range where both molecules have been measured,
i.e., between 1.5K and 3K. This justifies the merging of both data sets in order to obtain a larger temperature range.
Fitting Eq.~\ref{eq:kondowidth} to the merged data set yields a relatively poor fit as shown by the red line in
Fig.~\ref{fig:frotafits}(b), although somewhat better than the empirical expression $\Gamma_\emp(T)$ with fixed
temperature coefficient $\alpha=\pi$ (grey solid line), similar to the finding in Ref.~\onlinecite{Jacob:PRB:2023}.

We will now see that the mismatch with Eq.~\ref{eq:kondowidth} is caused by extrinsic broadening mechanisms. In STS,
the conductance spectra ($\didv$) are measured at finite temperatures $T$ where the Kondo peak is broadened, (i) due to
the \emph{intrinsic} temperature dependence of the Kondo peak~\cite{Nagaoka:PRL:2002,Osolin:PRB:2013,Jacob:PRB:2023},
and (ii) due to the presence of different \emph{extrinsic} broadening mechanisms in the STS
measurement~\cite{Gruber:JPCM:2018}. Thus, in order to obtain the actual \emph{intrinsic} halfwidth $\GammaK$ of the
Kondo peak and demonstrate Kondo behavior, one has to first remove the \emph{extrinsic} contributions from the measured
HWHM. Assuming a noise-optimized (electronic and mechanical) experimental setup, the two main sources of extrinsic
broadening are caused by Fermi-Dirac (FD) broadening of the tip and the voltage modulation for the lock-in detection.
A good fit with $\Gamma_\emp(T)$ can only be obtained by using $\alpha$ as a free fitting parameter (dashed grey line).

\begin{figure}
  \includegraphics[width=\linewidth]{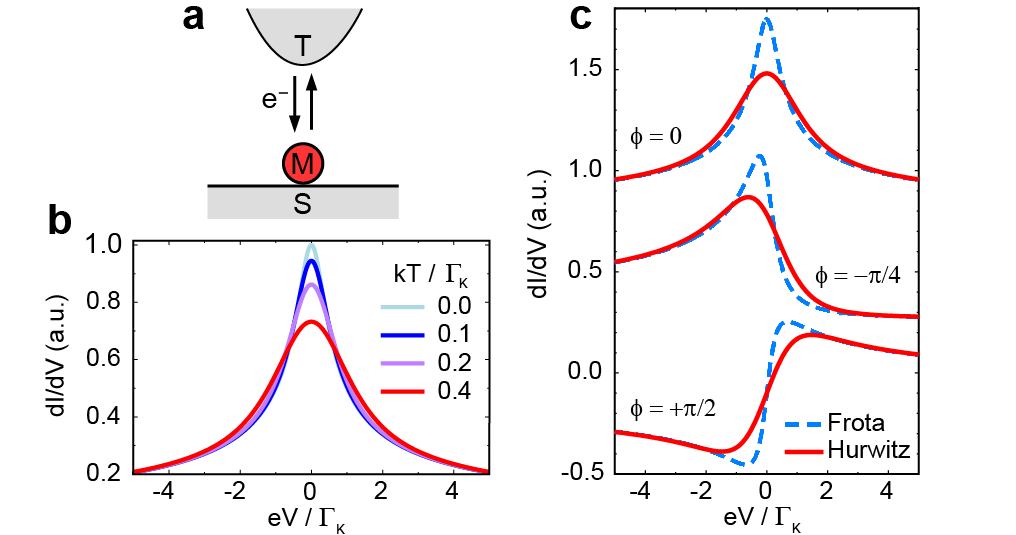}
  \caption{
    \label{fig:FD-smearing}
    (a) Schematic STM setup for measuring $\didv$ spectra of a molecule (M) on a surface (S) by an STM tip (T).
    (b) Simulated $\didv$ spectra assuming a Frota-Fano lineshape in the spectral function at
    different temperatures.
    (c) HF lineshapes (red solid lines) according to Eq.~\ref{eq:Hurwitz-Fano} at
    finite temperature $T=0.4\,\GammaK/\kB$ and corresponding Frota-Fano lineshapes (blue dashed lines)
    in underlying spectral function for different values of $\phi$ and fixed halfwidth $\GammaK=2.542\,\DeltaK$.
  }
\end{figure}

In order to properly incorporate the most important broadening mechanisms into the analysis
of the STS data we now resort to theory. 
The experimental situation of STS is depicted schematically in Fig.~\ref{fig:FD-smearing}(a):
A molecule (M) on a metallic substrate (S) is probed by an STM tip (T). Application of a
voltage $V$ to the sample drives a current $I$ from the sample electrode via the molecule to the
tip electrode.
Typically, in STS the coupling of M to T
is much weaker than the coupling of M to S.
In this situation, also called \emph{ideal STM limit}~\cite{Jacob:NL:2018},
M is effectively in equilibrium with S, and the $\didv$ of the current from the tip to
the molecule at the substrate is given by the convolution
\begin{equation}
  \label{eq:dIdV}
  G(V) \equiv \frac{dI}{dV} \propto \int d\omega \,  \left[-f^\prime(\omega)\right] \, A(\omega+eV) 
\end{equation}
where $f^\prime(\omega) = -\beta/\left(4\,{\rm cosh}^2(\beta\omega/2)\right)$ 
is the derivative of the tip's FD distribution, $\beta=1/\kB{T}$ with $T$ 
the temperature at the tip, and $A(\omega)$ the spectral function of the  
molecule~\cite{Note4}. We assume a fully thermalized system 
so that tip and sample have the same temperature $T$.

The convolution of the spectral function $A(\omega)$ with the FD derivative $f^\prime(\omega)$
leads to broadening of the Kondo resonance, to which we now simply refer as \emph{FD broadening}.
This is demonstrated in Fig.~\ref{fig:FD-smearing}(b) which shows the numerically calculated $\didv$
according to (\ref{eq:dIdV}) for different tip temperatures, assuming a Frota-Fano peak for the spectral
function, $A(\omega)\equiv{F(\omega)}$ with $\phi=0$ and constant $\Delta=\GammaK/2.542$.
Clearly, the effect is already considerable at temperatures of the order of $\TK\sim0.25\,\GammaK/\kB$.

In order to obtain the intrinsic linewidth $\Gamma=2.542\,\Delta$ of the Kondo resonance in the underlying
spectral function, while taking into account a finite temperature $T$ at the tip, 
we could simply fit (\ref{eq:dIdV}) to the experimental $\didv$ spectra. However, this 
requires to numerically evaluate the convolution (\ref{eq:dIdV})
several times during the fitting procedure, leading to considerable computational effort,
especially when lock-in modulation is also taken into account (see below).
Instead, we now derive an analytic expression for the lineshape of the Kondo resonance in the $\didv$
taking into account FD broadening, which allows to process the data very efficiently.
A lineshape for the Kondo peak 
at finite temperature in the spectral function was recently given in Ref.~\onlinecite{Jacob:PRB:2023}.
This lineshape reduces to a Frota-Fano lineshape in the $T\rightarrow0$ limit, but can also be well
approximated by a Frota-Fano lineshape at finite $T$~\cite{Osolin:PRB:2013}. We therefore describe
the Kondo resonance in the spectral function at some finite temperature $T$ by
the Frota-Fano lineshape, $A(\omega)\equiv{F(\omega)}$.
The Frota width parameter $\Delta$ is determined by the \emph{intrinsic} halfwidth of the
Kondo peak $\Gamma$ at temperature $T$ according to (\ref{eq:kondowidth}), i.e., $\Delta=\Gamma(T)/2.542$.
The resulting lineshape for the $\didv$ can be expressed analytically in terms of the
Hurwitz $\zeta$-function~\cite{supplemental}:
\begin{equation}
  \label{eq:Hurwitz-Fano}
  G(V) \propto \sqrt{\frac{\Delta}{8\tau} } \cdot \Re\left[
    e^{i\phi} \zeta\left( \frac{3}{2}, \frac{\Delta}{2\tau} + \frac{1}{2} +i\frac{eV}{2\tau} \right) 
\right]
\end{equation}
with the temperature parameter $\tau\equiv\pi \kB T$, as above.
The Hurwitz $\zeta$-function is a generalization of the Riemann $\zeta$-function, and
is defined by the infinite series $\zeta(s,a) = \sum_{n=0}^{\infty} 1/(n+a)^s$
where $s$ and $a$ can in general be complex, with $\Re[s]>1$ and $n+a\neq0$.
Fig.~\ref{fig:FD-smearing}(c) shows these \emph{Hurwitz-Fano} (HF) lineshapes according to (\ref{eq:Hurwitz-Fano}),
as red solid lines corresponding to three different Frota-Fano lineshapes (shown as blue dashed lines) in
the underlying spectral function $A(\omega)$ with different Fano phases $\phi$ as
indicated in the plots. 

\begin{figure}
  \includegraphics[width=\linewidth]{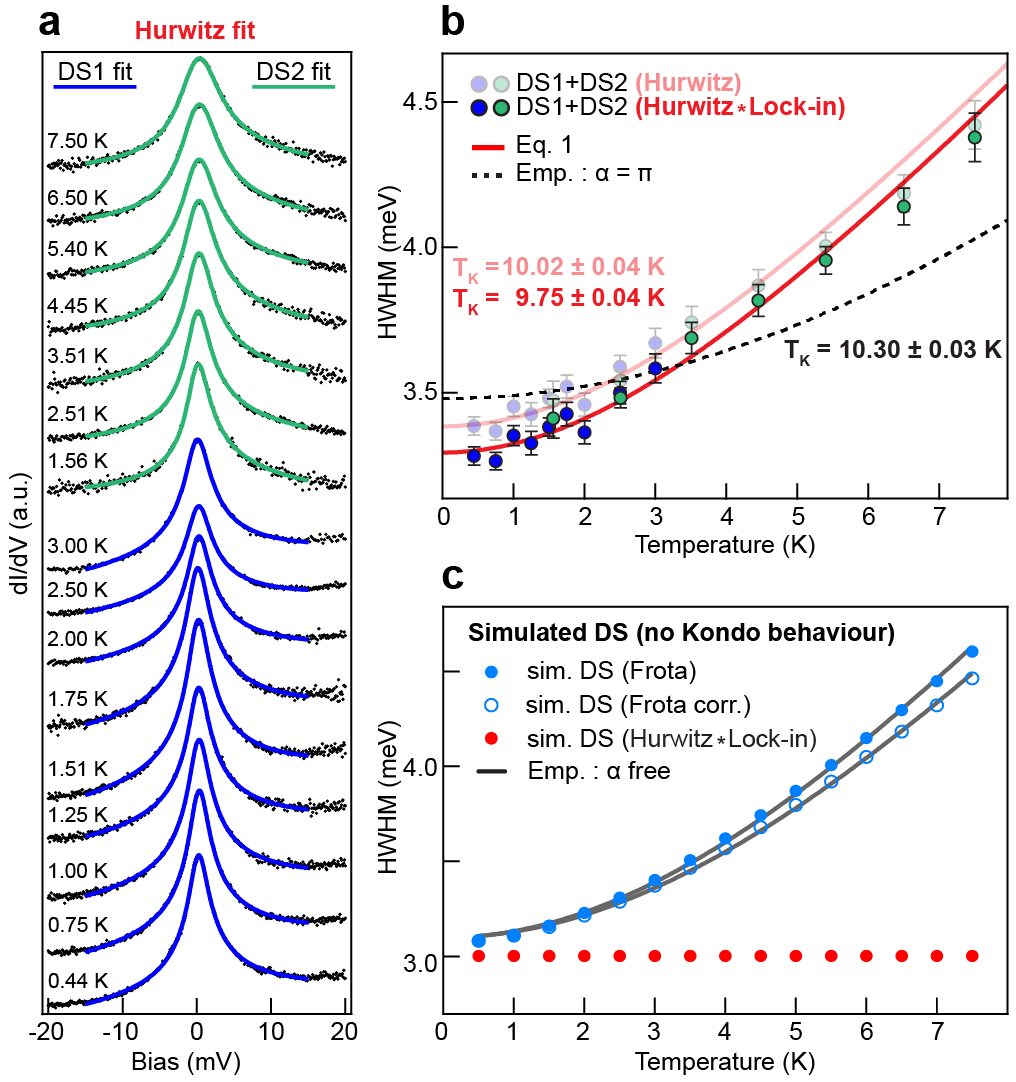}
  \caption{\label{fig:Hurwitz-fits}
    (a) $\didv$ spectra of DS1 and DS2 (black dots) fitted with HF lineshapes
    according to Eq.~\ref{eq:Hurwitz-Fano} (blue and green solid lines).
    (b) Intrinsic Kondo linewidths obtained from HF fits (transparent blue and green circles) 
    versus temperature, fitted with Eq.~\ref{eq:kondowidth} (transparent red line) 
    and with $\Gamma_\emp(T)$ with fixed $\alpha=\pi$ (black dashed line).
    Solid blue and green circles show intrinsic linewidths obtained from fitting with 
    Eq.~\ref{eq:lock-in}, taking into account both FD and lock-in broadening, 
    fitted with Eq.~\ref{eq:kondowidth} (solid red line).  
    $V_\mathrm{m}=0.5$\,mV for DS1 and $V_\mathrm{m}=0.4$\,mV for DS2. 
    Error bars show the estimated standard deviation.
    (c) Halfwidths versus temperature obtained from Frota (blue dots) and HF (red dots) fits 
    for simulated data of Frota peak with \emph{constant} intrinsic linewidth. 
    Also shown are temperature corrected Frota widths (blue circles) 
    using $\Gamma_{\rm corr.}$ (see text).
    Grey lines show fits with $\Gamma_\emp(T)$ resulting in $\alpha=5.32$ (Frota) and $\alpha=5.02$ (Frota corr.)
  }
\end{figure}

The HF lineshape (\ref{eq:Hurwitz-Fano}) can now be fitted~\cite{GitHub} directly to the Kondo resonances
in the finite-temperature $\didv$ spectra recorded by STS, as shown in Fig.~\ref{fig:Hurwitz-fits}(a).
We find that it fits slightly better the experimental data than the corresponding Frota fits~\cite{supplemental}.
Fitting the spectra with the HF lineshape yields smaller mean squared errors and the obtained width parameter
$\Delta$ is also more consistent with varying the fit range~\cite{supplemental}.
Importantly, the Frota width parameter $\Delta$ resulting from each fit now yields the  
\emph{intrinsic} halfwidth $\Gamma(T)=2.542\Delta$ of the Kondo resonance for a given temperature $T$.
As a result, the extracted halfwidths do follow the predicted intrinsic temperature broadening
and can be fitted well by Eq.~\ref{eq:kondowidth}, as shown by the transparent red line in
Fig.~\ref{fig:Hurwitz-fits}(b). This is in stark contrast to the halfwidths extracted from the
corresponding Frota fits in Fig.~\ref{fig:frotafits}(b).


Another important contribution to extrinsic broadening in STS experiments comes from the lock-in
modulation of the bias voltage. The effect on the spectra can be described by a further 
convolution of the differential conductance $G(V)$~\cite{Gruber:JPCM:2018,Esat:PRB:2015} as
\begin{equation}
    \label{eq:lock-in}
    \tilde{G}(V) = \int dV^\prime \, \chi_{\rm m}(V^\prime) \, G(V+V^\prime)
\end{equation}
where the lock-in function $\chi_{\rm m}$ is given by 
$\chi_{\rm m}(V)=2\sqrt{V_{\rm m}^2 -V^2}/\pi {V_{\rm m}}^2$ for $V\le{V_{\rm m}}$ and $\chi_{\rm m}(V)=0$ otherwise.
$V_{\rm m}=\sqrt2\,V_{\rm rms}$ is the amplitude of the bias modulation.
Thanks to the analytic solution (\ref{eq:Hurwitz-Fano}) of the first convolution (\ref{eq:dIdV}),
this second convolution can be computed numerically efficiently enough to be used in the fitting 
procedure~\cite{GitHub}. The resulting Frota width parameter $\Delta$ yields the Kondo width
without the extrinsic broadening due to FD smearing or lock-in modulation.
Thus, $\Gamma=2.542\Delta$ now essentially yields the \emph{intrinsic} halfwidth of the Kondo
resonance at a given temperature $T$ since the two most important STM inherent contributions of
extrinsic broadening have been removed.
Fig.~\ref{fig:Hurwitz-fits}(b) shows in  blue and green circles the intrinsic Kondo halfwidths
extracted in this way, which can be fitted even better by Eq.~\ref{eq:kondowidth}, resulting
in a mean square error of ${\rm MSE}_{\rm HL} = 1870\,\mu{\rm V}^2$ compared to
${\rm MSE}_{\rm H} = 2474\,\mu{\rm V}^2$ for the fits using just the HF lineshape.
In order to visualize the effect of FD broadening in our experimental dataset (DS1 and DS2),
in the Supplemental Material~\cite{supplemental} we compare the linewidths from Fig.~\ref{fig:frotafits}b
(Frota fit) with the linewidths from Fig.~\ref{eq:Hurwitz-Fano}b (Hurwitz fit), emphasizing the importance
of removing extrinsic FD broadening.  

So far we have shown the advantage of using the HF lineshape in order to directly obtain the intrinsic halfwidth of a
Kondo resonance, as well as the validity of Eq.~\ref{eq:kondowidth} to describe the correct temperature-dependent
broadening of the intrinsic halfwidth. The combination of these two expressions is crucial for the verification of a
Kondo resonance by its temperature dependence, as illustrated in Fig~\ref{fig:Hurwitz-fits}(c) for a simulated data set
where a Frota lineshape of constant halfwidth ($\Gamma = 3$\,mV) is broadened by FD and lock-in modulation
($V_\mathrm{m}=0.4$\,mV). Fitting this data set with a Frota-Fano lineshape (blue dots)~\cite{supplemental} yields an
increasing halfwidth, whose temperature dependence can be fitted well with the empirical expression $\Gamma_\emp(T)$,
thus giving the false impression of Kondo behavior. Also the often employed temperature correction of the linewidth
$\Gamma_{\mathrm{corr.}}=\sqrt{\Gamma^2-(1.75 \kB T)^2}$~\cite{Zhang:NComm:2013,Mishra:NNano:2019,Turco:JACSAu:2023} does
not properly compensate for the FD broadening, as shown by the blue circles in Fig~\ref{fig:Hurwitz-fits}(c).
Using the HF lineshape, on the other hand, yields the correct constant linewidth of the underlying peak (red circles),
revealing its non-Kondo nature. Similar results are obtained for a FD-broadened Lorentzian peak with a constant
linewidth~\cite{supplemental}. The empirical expression $\Gamma_\emp(T)$ is therefore not a suitable proof for Kondo
resonances and can only be used as an empirical way to extrapolate the Kondo temperature from such data sets if the
presence of the Kondo effect can already be assumed.

\begin{figure}
  \includegraphics[width=\linewidth]{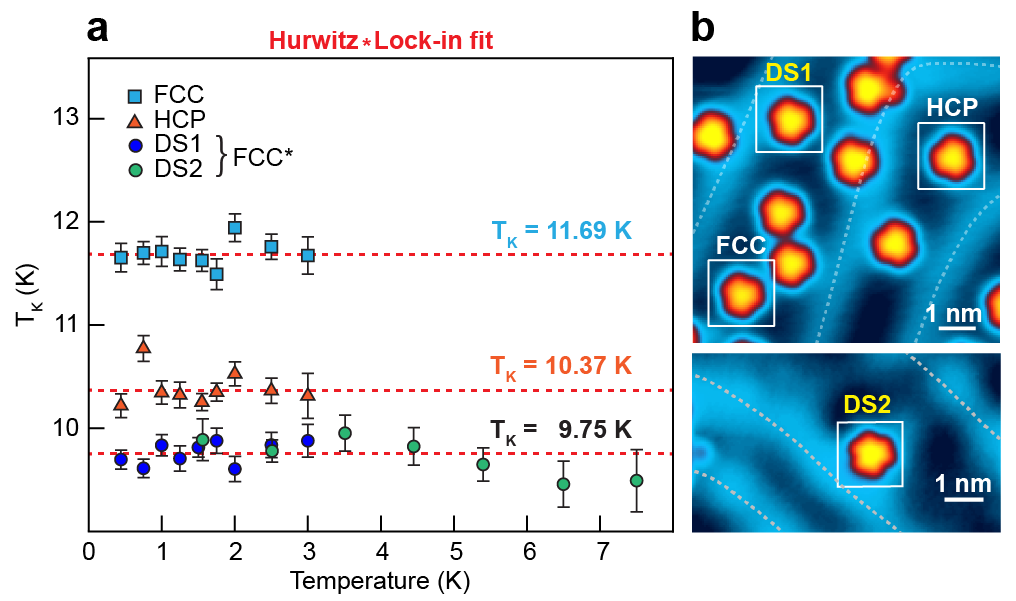}
  \caption{\label{fig:single-point-TK}
    (a) Kondo temperatures $\TK(T)$ determined for each $\didv$ spectrum individually by
    {HF$\ast$Lock-in} fits (Eq.~\ref{eq:lock-in}) and using the resulting intrinsic linewidth $\Gamma$
    in Eq.~\ref{eq:inverseKondo}. Error bars represent the estimated standard deviations.
    Red dashed lines indicate $\TK$ obtained by fitting the temperature dependence of the 
    Kondo linewidth for each adsorption site with Eq.~\ref{eq:kondowidth}~\cite{supplemental}.
    (b) STM images of phenalenyl molecules on Au(111) with boxes highlighting the investigated
    molecules. Scanning parameters for top image are $V = -2$\,V, $I = 100$\,pA and $V = -0.1$\,V,
    $I = 50$\,pA for the bottom image. Ridges of the herringbone reconstruction are marked with
    dashed white lines.
  }
\end{figure}

On the other hand, if no proof of Kondo behaviour is required, there is no need to perform temperature-dependent
measurements in order to determine $\TK$. According to (\ref{eq:kondowidth}) the intrinsic halfwidth $\Gamma$ at
a certain temperature depends only on $\DeltaK$ and the (known) experimental temperature. The relation can
therefore be rewritten as:
\begin{equation}
  \label{eq:inverseKondo}
  \DeltaK = \sqrt{\sqrt{\tfrac{1}{3}\Gamma^4+\tfrac{1}{3}\tau^2\Gamma^2
  +\alpha\tau^4} -\beta\tau^2-\gamma\Gamma^2}
\end{equation}
where $\alpha\equiv\left(2+\sqrt{3}\right)/6\sim0.622$, $\beta\equiv1-1/\sqrt{12}\sim0.711$ and $\gamma\equiv1-1/\sqrt3\sim0.423$
are constants and $\tau\equiv\pi \kB T$ the temperature parameter.
Using this relation, the Kondo temperature $\TK = \DeltaK/1.542\kB=\GammaK/3.92\kB$~\cite{Note1} can be determined accurately from
the intrinsic halfwidth of a single spectrum, taken at finite temperature. 
Fig.~\ref{fig:single-point-TK}(a) displays the $\TK$ obtained from the individual spectra of data sets DS1 and DS2 in dark blue and
green circles, respectively.
There is less than 3\% of variation between the individually determined Kondo temperatures and the $\TK$ obtained before by fitting
Eq.~\ref{eq:kondowidth} to the linewidths as a function of temperature. Similarly consistent Kondo temperatures are found by applying
this method to temperature-dependent data sets reported elsewhere~\cite{supplemental}.
The slight deviations from the expected constant behavior of the extracted $\TK$ versus temperature may be attributed to experimental
reasons, such as incomplete thermalization at higher temperatures~\cite{Gruber:JPCM:2018}.

We now apply the developed methodology to investigate the dependence of the Kondo temperature on the adsorption site of phenalenyl
on the herringbone reconstruction of the Au(111) surface. Along with data sets DS1 and DS2, acquired on two molecules in the
\textit{fcc} region of the Au(111) reconstruction close to the herringbone ridge (adsorption site labeled as FCC*), we also carried
out $\didv$ measurements for phenalenyl adsorbed in the middle of the \textit{fcc} (light blue) and \textit{hcp} (orange) regions.
By fitting each spectrum with a lock-in broadened HF lineshape and calculating $\TK$ using (\ref{eq:inverseKondo}), we find that
different adsorption geometries result in different Kondo temperatures, as clearly shown in Fig.~\ref{fig:single-point-TK}(a).
This observation can be rationalized by slight changes in the hybridization between molecule and metal substrate and the exponential
dependence of the Kondo temperature on the hybridization~\cite{Haldane:JPCSSP:1978}.
The robustness of the single-point $\TK$ determination, here shown for different adsorption sites and on an entire temperature series,
justifies the use of this method as a fast and reliable route to estimate $\TK$ of a Kondo system~\cite{Note2}.


To conclude, we have performed accurate STS measurements of a pure spin-1/2 Kondo system,
and developed tools for an efficient and accurate data analysis that properly take into account
extrinsic broadening of $\didv$ spectra in STS. A key element of the methodology is an analytic
expression for the Fermi-Dirac broadened Kondo lineshape in terms of the Hurwitz $\zeta$-function.
Fitting the spectra with this Hurwitz-Fano lineshape yields the intrinsic width of the Kondo
peak which fits very well with the recently derived expression (\ref{eq:kondowidth}) for the
Kondo width as a function of temperature, proving the validity of the theory~\cite{Jacob:PRB:2023}.
The procedure developed here allows to unequivocally prove Kondo behavior of a system probed by
STS~\cite{GitHub}. In contrast the established methodology of fitting the empirical
expression $\Gamma_\emp(T)$ to linewidths extracted from Frota fits to STS data may give a false impression of Kondo behavior.
Finally, our methodology also allows to obtain the intrinsic Kondo width
at $T=0$ and corresponding Kondo temperature $\TK$ from a \emph{single} spectrum at finite
temperature.

\begin{acknowledgments}
  PL, MA, RD and SCG acknowledge funding by the European Research Council (ERC-2017-AdG no.~788185 ``Artificial Designer Materials'')
  and the Academy of Finland (Academy professor funding nos. 318995 and 320555 and Academy research fellowships 347266 and 353839).
  DJ and NS acknowledge financial support via Grant PID2020-112811GB-I00 from MCIN/AEI/10.13039/501100011033 and Grant No. IT1453-22
  from the Basque Government. This research made use of the Aalto Nanomicroscopy Center (Aalto NMC) facilities.
  MJ and AB acknowledge funding by the Swiss National Science Foundation (PP00P2\_170534, PP00P2\_198900, TMCG-2\_213829) and the
  European Union’s Horizon 2020 research and innovation programme (ERC Starting grant INSPIRAL, Grant No. 716139).
  MJ, AB, RF, PR, ET and NK acknowledge funding by the Swiss National Science Foundation under Grant no. CRSII5\_205987.
  RF, PR, ET and NK further acknowledge funding by the European Union's Horizon 2020 research and innovation programme under the
  Marie Skłodowska--Curie grant no. 813036 and CarboQuant funded by the Werner Siemens Foundation.
  For the purpose of Open Access, the author has applied a CC BY public copyright license to any Author Accepted Manuscript version arising from this submission.
\end{acknowledgments}

E.T. and M.A. contributed equally to this work.

\bibliography{refs,footnotes}

\end{document}


\title{Supplemental Material for:\\
Demonstrating Kondo behavior by temperature-dependent scanning tunneling spectroscopy
}

\author{Elia Turco}
\email{elia.turco@empa.ch}
\affiliation{nanotech@surfaces Laboratory, Empa - Swiss Federal Laboratories for Materials Science and Technology, 8600 Dübendorf, Switzerland}

\author{Markus Aapro}
\email{markus.aapro@aalto.fi}
\affiliation{Department of Applied Physics, Aalto University, 00076, Aalto, Finland}

\author{Somesh C. Ganguli}
\affiliation{Department of Applied Physics, Aalto University, 00076, Aalto, Finland}

\author{Nils Krane}
\affiliation{nanotech@surfaces Laboratory, Empa - Swiss Federal Laboratories for Materials Science and Technology, 8600 Dübendorf, Switzerland}

\author{Robert Drost}
\affiliation{Department of Applied Physics, Aalto University, 00076, Aalto, Finland}

\author{Nahual Sobrino}
\affiliation{Departamento de Polímeros y Materiales Avanzados: Física, Química y Tecnología,
Universidad del País Vasco UPV/EHU, Av. Tolosa 72, E-20018 San Sebastián, Spain}

\author{Annika Bernhardt}
\author{Michal Juríček}
\affiliation{Department of Chemistry, University of Zurich, Winterthurerstrasse 190, 8057 Zurich, Switzerland}

\author{Roman Fasel}
\affiliation{nanotech@surfaces Laboratory, Empa - Swiss Federal Laboratories for Materials Science and Technology, 8600 Dübendorf, Switzerland}
\affiliation{Department of Chemistry, Biochemistry and Pharmaceutical Sciences, University of Bern, 3012 Bern, Switzerland}

\author{Pascal Ruffieux}
\affiliation{nanotech@surfaces Laboratory, Empa - Swiss Federal Laboratories for Materials Science and Technology, 8600 Dübendorf, Switzerland}
\email{pascal.ruffieux@empa.ch}

\author{Peter Liljeroth}
\affiliation{Department of Applied Physics, Aalto University, 00076, Aalto, Finland}
\email{peter.liljeroth@aalto.fi}

\author{David Jacob}
\email{david.jacob@ehu.es}
\affiliation{Departamento de Polímeros y Materiales Avanzados: Física, Química y Tecnología,
Universidad del País Vasco UPV/EHU, Av. Tolosa 72, E-20018 San Sebastián, Spain}
\affiliation{IKERBASQUE, Basque Foundation for Science, Plaza Euskadi 5, E-48009 Bilbao, Spain}

\maketitle

\section{Methods}

The temperature-dependent measurements in this work were carried out in an Unisoku USM-1300 STM with a 320\,mK base sample temperature.
The hydrogen-passivated phenalenyl precursors were deposited via flash deposition from a silicon wafer onto a clean Au(111) surface held at room temperature.
The precursor molecule synthesis and dehydrogenation by atomic manipulation into the target spin-1/2 nanographene is described in Ref.~\onlinecite{Turco:JACSAu:2023}. Nanonis electronics were used for STM control and data collection. 
The tunneling current signal was amplified with a variable-gain FEMTO DLPCA-200 preamplifier and low-pass filters were used in all electronic inputs to the STM. 
STS spectra were measured with standard lock-in detection using modulation voltages of $V_{\rm m} = 400$\,$\mu$V (data set DS2) and $500$\,$\mu$V (DS1, HCP and FCC) and, if not stated otherwise, the Kondo resonances in the spectra were fitted with a fit range of $\pm15$\,mV. Data sets DS1, HCP and FCC were measured with the same tip apex, whereas DS2 was measured with a different tip.
The data was processed with Wavemetrics Igor Pro software, whereas the fit procedures were implemented both in Python and Igor programming language. The fit procedures are open-source and will be made available on GitHub~\cite{github}.

\newpage

\section{Fitting of DS1 and DS2 with Frota-Fano Lineshape}

\begin{figure}[h!]
  \includegraphics[width=0.9\linewidth]{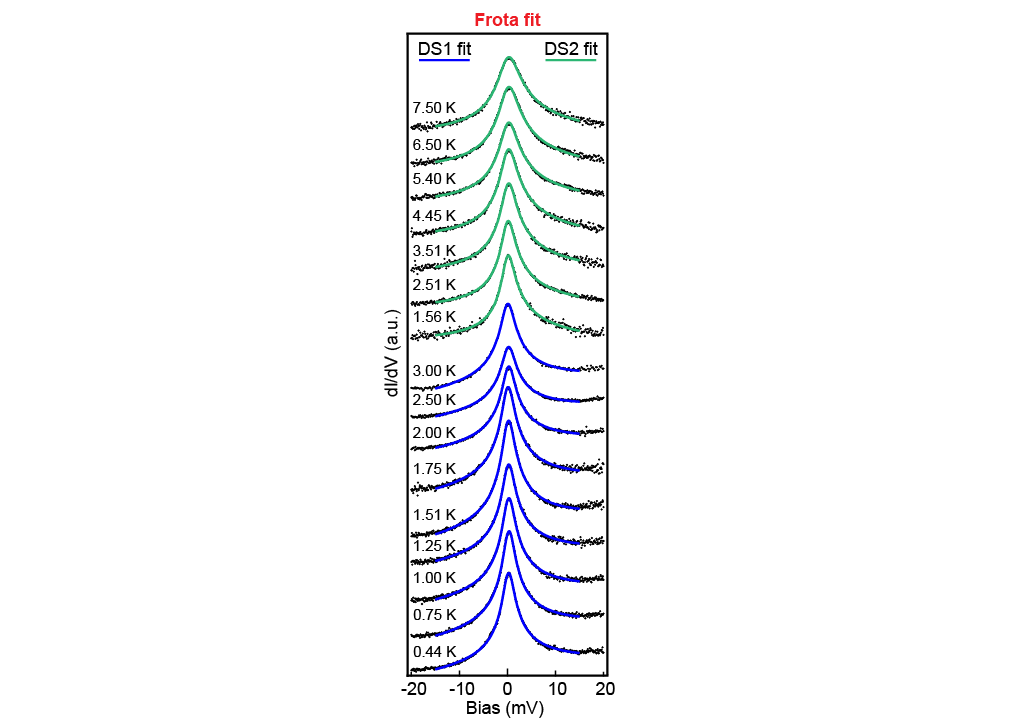}
  \caption{\label{fig:FrotaFits}
    $\didv$ spectra of merged data sets DS1 and DS2 for similarly adsorbed molecules together with the respective Frota fit (blue [DS1] and green [DS2] solid lines). Open feedback parameters: $V = -25$ mV, $I$ = 500 pA; Lock-in modulation: DS1 $V_{\rm m}$ = 0.5\,mV; DS2 $V_{\rm m}$ = 0.4\,mV. 
    }
\end{figure}

\newpage

\section{Derivation of the Hurwitz-Fano lineshape}
\begin{SCfigure}[][b]
	\centering
	\includegraphics[width=0.25\textwidth]{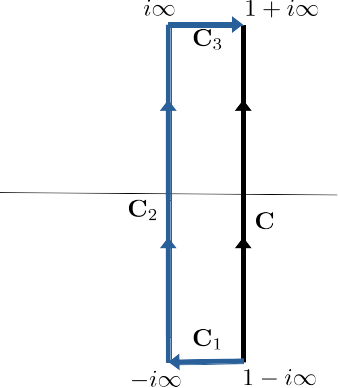}
	\caption{Illustration of complex contours. The black line, labeled as $\mathbf{C}$, represents a direct path from $1-i\infty$ to $1+i\infty$. The blue contour comprises three segments: $\mathbf{C}_1$ begins at $1-i\infty$  and moves leftwards to reach the imaginary axis at $-i\infty$ , $\mathbf{C}_2$ ascends along the imaginary axis to $i\infty$ , and $\mathbf{C}_3$ proceeds rightwards to conclude at $1+i\infty$.}
	\label{fig1}
\end{SCfigure}

In this section we calculate the integral of Eq~(2) in the main text assuming a Frota peak for the spectral function $A(\omega)\equiv F(\omega)$. The integral can be written as
\begin{align}
	G  = \int_{-\infty}^{\infty}[-f'(\omega)]A(\omega+eV)d\omega= \frac{\beta}{4}\text{Re}\left[e^{i\phi}\int_{-\infty}^{\infty}\frac{\text{sech}^{2}\left(\frac{\beta}{2}(\omega-eV)\right)}{\sqrt{1+i\omega/\Delta}}d\omega\right],
\end{align}
where $\beta = 1/\kB T$. Choosing $z=1+i\omega/\Delta$ and defining $a=\frac{\Delta\beta}{2}$, the integral can be rewritten as
\begin{align}
	G = \frac{1}{2}\text{Re}\left[-iae^{i\phi}\int_{1-i\infty}^{1+i\infty}\frac{\text{sech}^2(-ia(z-1)-aeV/\Delta)}{\sqrt{z}}dz\right]
 	\label{int_aux}
\end{align}
along $\mathbf{C}$, see Fig.~(\ref{fig1}). Since the integrand is analytic within the region enclosed by the contour $\mathbf{C}$ and the composite contour   $\mathbf{C}_1+\mathbf{C}_2+\mathbf{C}_3$, Eq.~(\ref{int_aux}) is equal to
\begin{align}
	G =\frac{-a}{2}\lim_{k\to\infty}\left(\text{Re}\left[ie^{i\phi}\left(\int_{1-ik}^{-ik}g(z)dz+\int_{-ik}^{ik}g(z)dz\right.\right.\right.\left.\left.\left.+\int_{ik}^{1+ik}g(z)dz\right)\right]\right)
	\label{int_aux0}
\end{align}
with $g(z) = \text{sech}^2(-ia(z-1)-aeV/\Delta)z^{-\frac{1}{2}}$.  The first and third integrals of Eq.~(\ref{int_aux0}) vanish due to the exponential decay of $\text{sech}(z)$ for large real values of $z$ when $k\to\infty$, given their contours of integration.  The finite contribution can be written again as a real integral making the change $x=-iaz$
\begin{align}
	I_1=\int_{-\infty}^{\infty}\frac{\text{sech}^2(x+ia-aeV/\Delta)}{\sqrt{ix/a}}dx
	\label{int_aux1}
\end{align}
with $G=\text{Re}(e^{i\phi}I_1)/2$.

The contribution along negative $x$ can be expressed as 
\begin{align}
	\int_{-\infty}^{0}\frac{\text{sech}^2(x+ia-aeV/\Delta)}{\sqrt{ix/a}}dx=\sqrt{a}e^{i\pi/4}\int_{0}^{\infty}\frac{\text{sech}^2(x-ia+aeV/\Delta)}{\sqrt{x}}dx
	\label{int_aux2}
\end{align}
where it has been used that  $\text{sech}(-z)=\text{sech}(z)$ and $e^{\mp i\pi/4}=(\pm i)^{-1/2}$ . Therefore
\begin{align}
	G= \frac{\sqrt{a}}{2}\text{Re}\left(e^{i(\phi+\pi/4)}\int_{0}^{\infty}\frac{\text{sech}^2(x+d_+)}{\sqrt{x}}dx+e^{i(\phi-\pi/4)}\int_{0}^{\infty}\frac{\text{sech}^2(x+d_-)}{\sqrt{x}}dx \right)
	\label{int_aux3}
\end{align}
with $d_{\pm}=\mp ia\pm aeV/\Delta$. On the other hand, since 
$
\text{sech}^2(x+d_{\pm})=-4\frac{\partial}{\partial x}  \left([1+e^{2x+2d_{\pm}}]^{-1}\right)
$
one finds that
\begin{align}
	G =-2\sqrt{a}~\text{Re}\left[e^{i(\phi+\pi/4)}\int_{0}^{\infty}x^{-1/2}\frac{\partial}{\partial x}  \left(\frac{1}{1+e^{2x+2d_+}}\right)dx+e^{i(\phi-\pi/4)}\int_{0}^{\infty}x^{-1/2}\frac{\partial}{\partial x}  \left(\frac{1}{1+e^{2x+2d_-}}\right)dx\right].
\end{align}
Making use of the well-known complete Fermi-Dirac integral \cite{gradshteyn2014table,dingle1957fermi}
\begin{align}
	F_j(t) = \frac{2^{j+1}}{\Gamma(j+1)}\int_{0}^{\infty}\frac{x^{j}}{e^{2x-t}+1}dx=-\text{Li}_{j+1}(-e^{t})
\end{align}
where $\text{Li}_s(z)$ is the polylogarithm function, identifying $j=-1/2$, $t=-2d_{\pm}$, and taking into account that $\frac{\partial}{\partial x}  \left[1+e^{2x-t}\right]^{-1} = -2\frac{\partial}{\partial t}  \left[1+e^{2x-t}\right]^{-1}$, we can express the integral as 
\begin{align}
	G =\sqrt{2a\pi}~\text{Re}\left[e^{i(\phi+\pi/4)}\left.\frac{dF_{-1/2}(t_0)}{dt_0}\right|_{t_0=-2d_+}+e^{i(\phi-\pi/4)}\left.\frac{dF_{-1/2}(t_0)}{dt_0}\right|_{t_0=-2d_-}\right]
\end{align}
and using  $\frac{dF_{j}(t)}{dt}=F_{j-1}(t)$, we obtain that 
\begin{align}
	G &=\sqrt{2a\pi}~\text{Re}\left[e^{i(\phi+\pi/4)}F_{-3/2}(-2d_+) +e^{i(\phi-\pi/4)}F_{-3/2}(-2d_-)\right]
	\nonumber\\&=
	-\sqrt{\frac{\pi\Delta}{ T}}~\text{Re}\left[e^{i(\phi+\pi/4)}\text{Li}_{-1/2}(-e^{i\beta\Delta-\beta eV})+e^{i(\phi-\pi/4)}\text{Li}_{-1/2}(-e^{-i\beta\Delta+\beta eV})\right].
\end{align}
Making use of the relation between the polylogarithm and the Hurwitz zeta function
\begin{equation}
	\text{Li}_\nu(z) = (2\pi)^{\nu - 1} i \Gamma(1 - \nu) \left[e^{-\frac{\pi i \nu}{2}} \zeta\left(1 - \nu, \frac{\log(-z)}{2\pi i}+\frac{1}{2}\right) - e^{\frac{\pi i \nu}{2}}\zeta\left(1 - \nu,- \frac{\log(-z)}{2\pi i}+\frac{1}{2}\right)\right],
\end{equation}
we can finally write
\begin{align}
	G =\sqrt{\frac{\beta\Delta}{8\pi}}\text{Re}\left[e^{i\phi}\zeta{\left(\frac{3}{2},\frac{1}{2}+\frac{\beta\Delta}{2\pi}+i\frac{\beta eV}{2\pi}\right)} \right].
\end{align}

\newpage

\section{Definitions of the Kondo temperature}

In the literature many different definitions of the Kondo temperature can be found.
In this work we are using Wilson's thermodynamic definition $\TK$~\cite{Wilson:RMP:1975}, 
corrected by Wiegman and Tsvelick~\cite{Wiegmann:JPC:1983}, in terms of the magnetic 
susceptibility $\chi_{\rm imp}(T)=0.68(g\mu_B)^2/4\kB(T+\sqrt2\TK)$ in the crossover regime ($0.5\TK<T<16\TK$, see Eq.~4.60 in Ref.~\onlinecite{Hewson:book:1997}):
\begin{equation}
    \label{eq:Def_TK_Wilson}
    \chi_{\rm imp}(\TK) = \frac{0.0704\,(g\mu_B)^2}{\kB\TK} 
\end{equation}
where $g$ is the electronic g-factor, and $\mu_{\rm B}$ the Bohr magneton. 
On the other hand, in the low-temperature strong-coupling regime ($T\ll\TK$), 
the magnetic susceptibility becomes constant: $\chi_{\rm imp}(T) = w\,(g\mu_B)^2/4\kB\TK$ 
where $w$ is Wilson's number $w=0.4128$ (see Eq.~4.58 in Ref.~\onlinecite{Hewson:book:1997}).
In this regime one often uses the strong-coupling definition of the Kondo 
temperature $T_0$ (sometimes also denoted by $\TL$) instead of $\TK$ which 
is related to Wilson's $\TK$ by $\TK=w\,T_0$.
Thus the magnetic susceptibility in the strong-coupling regimes becomes:
\begin{equation}
    \label{eq:Def_T0}
    \chi_{\rm imp}(T) = \frac{(g\mu_B)^2}{4\kB T_0}
\end{equation}
In Ref.~\onlinecite{Jacob:PRB:2023} Wilson's $\TK$ is related to the intrinsic $T=0$ halfwidth $\GammaK$ of the Kondo peak by $\GammaK={2\cdot2.542\kB\TK/\pi{w}} = {3.92\,\kB\TK}$, and hence $\GammaK={2\cdot2.542\,\kB{T_0}/\pi}={1.618\,\kB T_0}$.

In experimental work the Kondo temperature is often defined by the empirical formula
used to fit the temperature-dependent broadening of the Kondo resonance, where one of
two definitions are most often used:
First, in the equation for the temperature-dependent width of the Kondo peak originally 
derived from Fermi liquid theory by Nagaoka {\it et al.} in Ref.~\onlinecite{Nagaoka:PRL:2002} 
\begin{equation}
  \label{eq:nagaoka}
  \Gamma_{\rm emp}(T) = \sqrt{ (\alpha \kB T)^2+2(\kB T_\mathrm{K,N})^2},
\end{equation}
with $\alpha$ being either fixed ($\alpha=\pi$) or used as fitting parameter,
yields $\GammaK={\sqrt{2}\kB T_\mathrm{K,N}}$.
Second, a variation of that equation used in several experimental works~\cite{Zhang:NComm:2013,Khajetoorians:NNano:2015,Mishra:NNano:2019,Li:PRL:2020,Turco:JACSAu:2023}
is given by
\begin{equation}
  \label{eq:zhang}
  \Gamma_{\rm emp}(T) = \frac{1}{2}\sqrt{ (\alpha \kB T)^2+(2\kB T_\mathrm{emp})^2}.
\end{equation}
and thus yields $\GammaK=\kB T_\mathrm{emp}$, i.e. the Kondo temperature in that definition is identical with the
intrinsic width $\GammaK$ of the Kondo peak at $T=0$.
Table~\ref{tab:TK2} displays the different definitions of the Kondo temperature and their numerical conversion factors, including the 
halfwidth $\GammaK$ of the Kondo peak.

\begin{table}[h]
\centering
    \begin{tabular*}{0.7\linewidth}{r@{\hspace{3ex}}||@{\extracolsep\fill}cccc}
                          & $\times\,\kB \TK$ & $\times\,\kB T_0$ & $\times\,\kB T_{\rm K,N}$ &  $\times\,\GammaK$ \\[1ex]
                          \hline
        $\kB\TK=$         & 1      & 0.4128 & 0.3608 & 0.2551 \\[1ex]
        $\kB T_0=$        & 2.422  & 1      & 0.8740 & 0.6180 \\[1ex]
        $\kB T_{\rm K,N}=$  & 2.772  & 1.144  & 1      & 0.7071 \\[1ex]
        $\GammaK=$        & 3.920  & 1.618  & 1.414  & 1      
    \end{tabular*}        
\caption{\label{tab:TK2} 
Numerical conversion factors between different definitions of the Kondo temperature, where
$\TK$ is Wilson's definition, $T_0$ the strong-coupling definition (sometimes denoted $\TL$),
$T_\mathrm{K,N}$ the definition introduced by Nagaoka {\it et al.}~\cite{Nagaoka:PRL:2002},
$\GammaK$ is the halfwidth of the Kondo peak and ${\kB=8.6173\cdot10^{-2}\meV/{\rm K}}$ is 
Boltzmann's constant. The quantities given in the leftmost column can be expressed by the 
quantities given in the top row multiplied by the conversion factor given by the corresponding 
table entry, e.g., $\kB\TK = 0.2551\times\GammaK$.
}
\end{table}

\newpage

\section{Comparison of Frota and Hurwitz fits of data set DS2}

\begin{figure}[h!]
  \includegraphics[width=\linewidth]{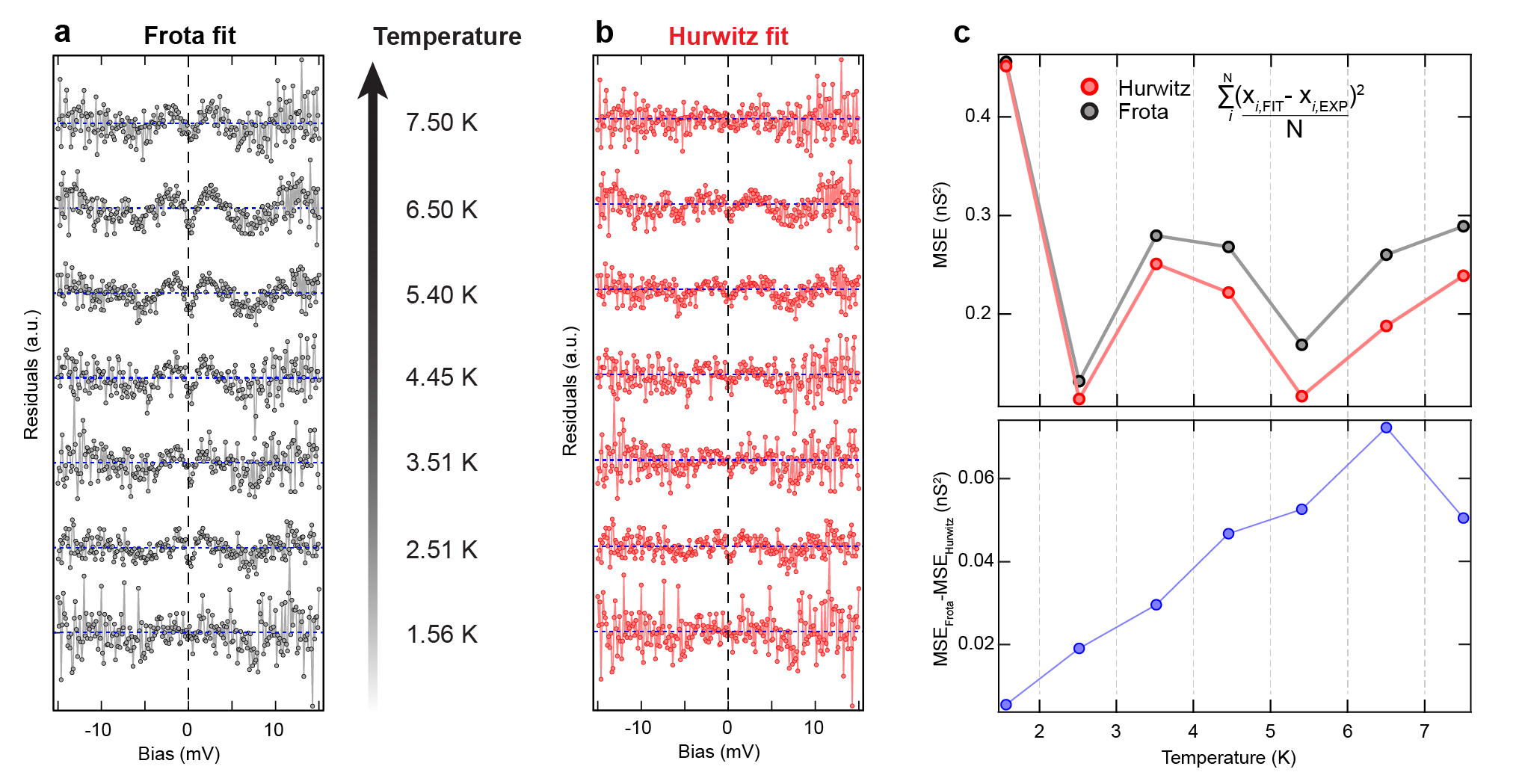}
  \caption{\label{fig:residuals}
    Residuals analysis of Frota and Hurwitz fitting of DS2 spectra.
    (a,b) Residuals of Frota and Hurwitz fits, respectively, of the spectra acquired in the temperature range 1.56\,K - 7.50\,K. (c) The MSE is plotted in the top graph as a function of the temperature for both Frota and Hurwitz fits, while the bottom graph shows their difference as a function of temperature.  
    }
\end{figure}

The analysis reported in Fig. \ref{fig:residuals} essentially reveals that the Hurwitz-Fano lineshape fits overall better than the Frota-Fano function, here shown for data set DS2. Panel (c) shows that the mean squared error (MSE), which is a value representing the discrepancy of the fit with respect to the experimental data, is always smaller for the Hurwitz fit than the Frota fit, and that this difference becomes larger for higher temperatures. Notably the $\didv$ spectrum acquired at $1.56$K, despite being the lowest temperature of data set DS2, was affected by an overall higher noise level, which explains its high MSE value. Panels (a,b) reveal the bias dependency of the residuals, clearly showing that the Frota function fails to fit the raw experimental data in the close proximity of the Fermi level.    

\newpage

\section{Fit range dependence of extracted halfwidths}

\begin{figure}[h!]
  \includegraphics[width=0.7\linewidth]{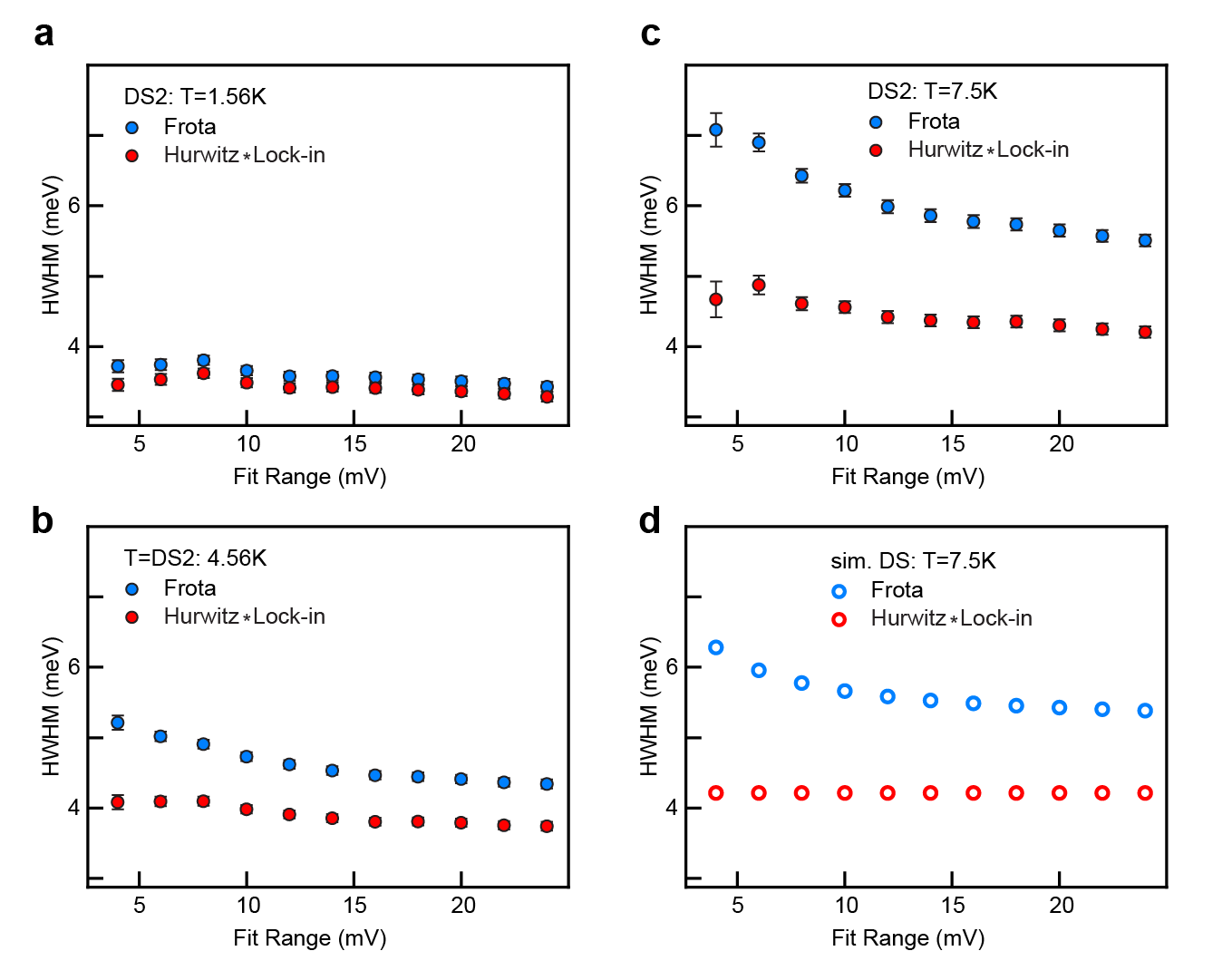}
  \caption{\label{fig:FittingRange}
  (a-c) Halfwidth of three $\didv$ spectra from data set DS2, determined by fitting a Frota-Fano (blue) or lock-in-convoluted Hurwitz-Fano (red) lineshape using different fit ranges. The error bars represent the estimated standard deviation.
  (d) Fit-range dependent halfwidth of a simulated Kondo spectrum ($\Gamma_K = 3.0\,\mathrm{meV}$, $V_\mathrm{m} = 0.4\,\mathrm{mV}$), obtained by Frota (red) and Hurwitz$\ast$Lock-in (blue) fit.
    }
\end{figure}

Since the Hurwitz-Fano function describes the lineshape of a $\didv$ spectrum measured at finite temperature better than the Frota-Fano function, it is also less dependent on the used fit range. In Fig.~\ref{fig:FittingRange}(a-c) three spectra from data set DS2 were fitted with different fit ranges using both Frota-Fano, as well as lock-in-convoluted Hurwitz-Fano lineshapes. At low temperatures (Fig.~\ref{fig:FittingRange}a) both fits yield similar and consistent results, whereas at higher temperatures (Fig.~\ref{fig:FittingRange}b,c) the Frota fit overestimates the halfwidth, especially for small fit ranges. The Hurwitz fit on the other hand yields a much more constant halfwidth. The small increase of the halfwidth obtained by Hurwitz-Fano at smaller fit ranges ($< 10$\,mV) can be attributed to a small contribution of noise broadening.

Fig.~\ref{fig:FittingRange}(d) displays the halfwidths obtained by Frota and Hurwitz fits applied to a simulated data set of a Kondo resonance with $\Gamma_K = 3.0\,\mathrm{meV}$ and $V_\mathrm{m} = 0.4\,\mathrm{mV}$. Similar to the experimental results, the Frota fit yields a strong overestimation of the halfwidth and varies strongly with the fit range, whereas the Hurwitz fit yields constant halfwidth.
Please note that the obtained halfwidth by the Hurwitz fit is not $\Gamma_K$ but $\Gamma$ according to Eq.~(1) of the main text.

\newpage

\section{Fits of Simulated Data without Kondo behaviour}

\subsection{Frota lineshape}

\begin{figure}[h!]
  \includegraphics[width=0.5\linewidth]{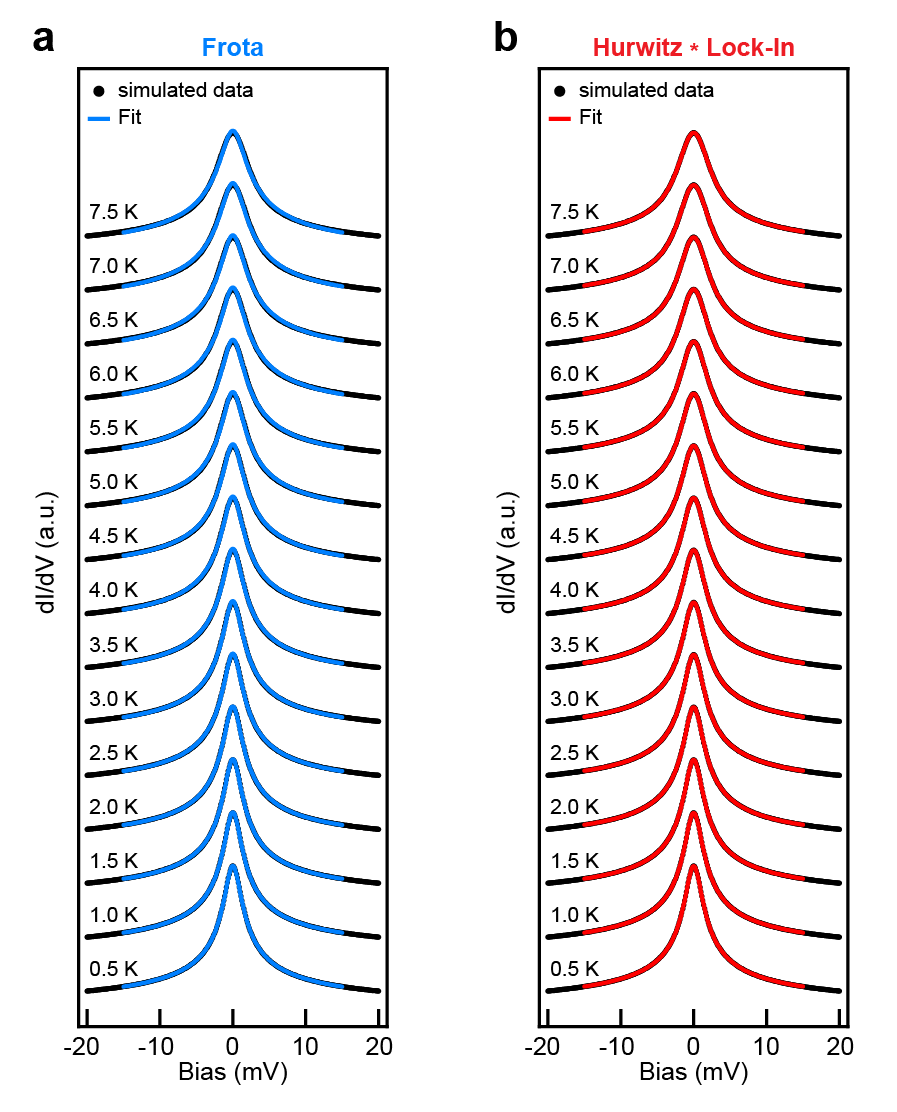}
  \caption{\label{fig:simulatedDataFitting}
    Simulated temperature-dependent $\didv$ spectra of a Frota-Fano lineshape with constant \emph{intrinsic} halfwidth (black lines). Blue lines in (a) and red lines in (b) display Frota and Hurwitz$\ast$Lock-in fits, respectively.
    }
\end{figure}

We simulated a temperature-dependent $\didv$ data set of a resonance at zero bias voltage with Frota-Fano lineshape. The \emph{intrinsic} linewidth ($\Gamma=3$\,mV) of the resonance does not change with temperature and is only broadened by Fermi-Dirac smearing and the applied voltage modulation ($V_{\rm m}=0.4$\,mV). Fig.~\ref{fig:simulatedDataFitting} displays both the Frota and the Hurwitz$\ast$Lock-in fits to the simulated data set. The obtained halfwidths are displayed in the main text.

\newpage

\subsection{Lorentzian lineshape}
If we consider a typical non-Kondo lineshape, i.e. a Lorentzian, and add, both, the Fermi-Dirac broadening and voltage modulation ($V_{\rm m}=0.4$\,mV), we obtain the simulated data reported below. 
If we fit the spectra with a Hurwitz$\ast$Lock-in lineshape, the obtained halfwidth will not yield a constant value (as expected). Instead it even even decreases at higher temperature. Therefore it does not follow the expected temperature evolution of a Kondo peak as described by eq. (1), and it is obvious that the zero-bias resonance is not a Kondo peak.
A fit with a Frota-Fano lineshape on the other hand, does result in an increase of the halfwidth with temperature and can therefore lead to a misattribution as a Kondo peak, even when the often used temperature correction term $\Gamma_\mathrm{corr.}=\sqrt{\Gamma^2-(1.75 \kB T)^2}$ is used.

\begin{figure}[h!]
  \includegraphics[width=\linewidth]{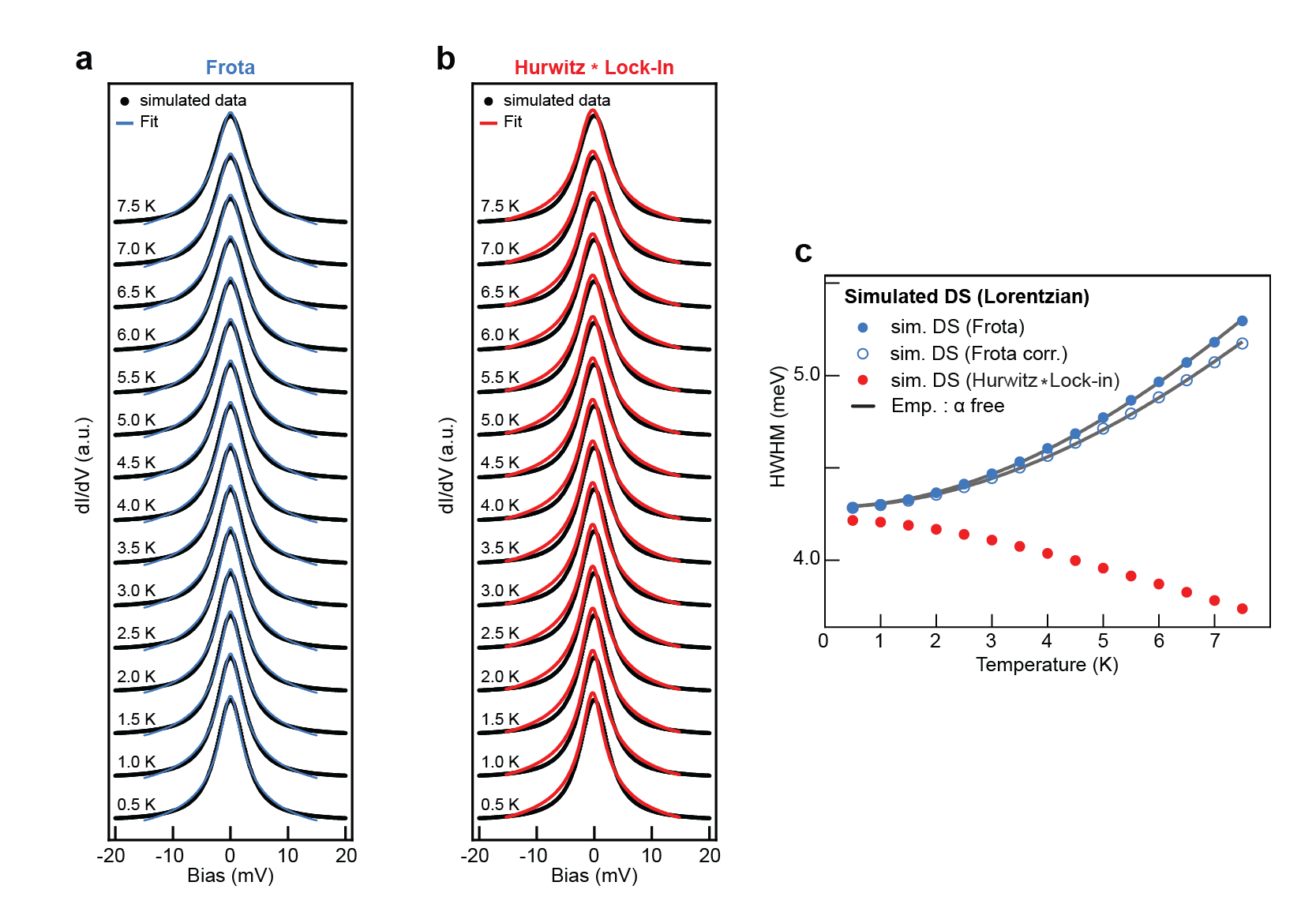}
  \caption{ 
    \label{fig:simFit_lorentzian}
    Simulated temperature-dependent $\didv$ spectra of a Lorentzian lineshape with constant \emph{intrinsic} halfwidth (black lines). Blue lines in (a) and red lines in (b) display Frota and Hurwitz$\ast$Lock-in fits, respectively. 
    (c) Halfwidths versus temperature obtained from Frota (blue dots) and Hurwitz (red dots) fits to a simulated data set without intrinsic broadening, as well as the often used temperature correction term $\Gamma_\mathrm{corr.}=\sqrt{\Gamma^2-(1.75 \kB T)^2}$ for the Frota halfwidth (blue circles).
    Grey lines show fit with $\Gamma_\mathrm{emp}(T)$ resulting in $\alpha=4.83$ (Frota) and $\alpha=4.51$ (Frota corr.)
    }
\end{figure}

\newpage
\section{The importance of removing Fermi-Dirac (FD) broadening}

We compare the Kondo linewidths obtained by fitting the joint (DS1+DS2) dataset with a Frota function and Hurwitz function as well as the often used temperature correction $\Gamma_\mathrm{corr.}=\sqrt{\Gamma^2-(1.75 \kB T)^2}$. It is clear that Fermi-Dirac broadening has a significant influence already for temperatures $T > 2K$ and that the square root temperature correction does not effectively remove the FD broadening.  

\begin{figure}[h!]
  \includegraphics[width=\linewidth]{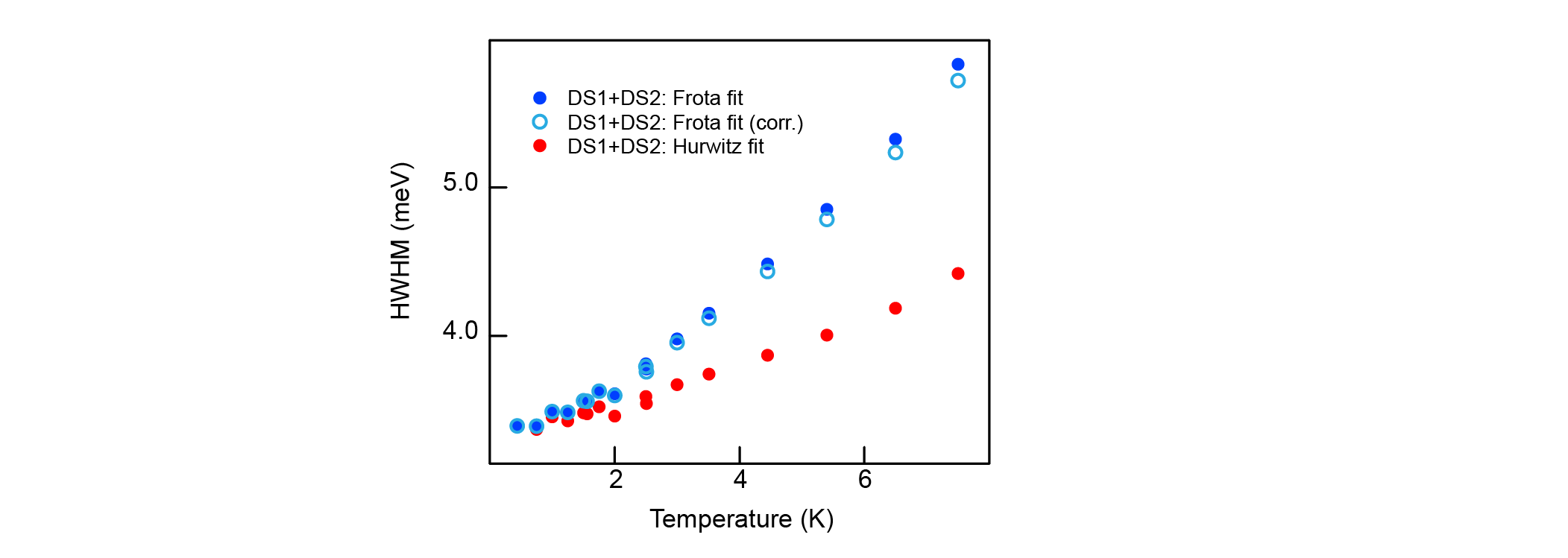}
  \caption{ 
    \label{fig:FD effect}
     Kondo linewidths obtained by fitting the joint (DS1+DS2) dataset with a Frota function (in red) and Hurwitz function (in blue) as well as the often used temperature correction term $\Gamma_\mathrm{corr.}=\sqrt{\Gamma^2-(1.75 \kB T)^2}$ for the Frota halfwidth (blue circles). The Figure highlights the importance of removing the Fermi-Dirac broadening, which, for this specific system, has a significant effect for $T \geq 2$ K.
}
\end{figure}
\newpage

\section{Temperature-dependent $\didv$ spectroscopy of FCC and HCP molecules }

\begin{figure}[h!]
  \includegraphics[width=0.7\linewidth]{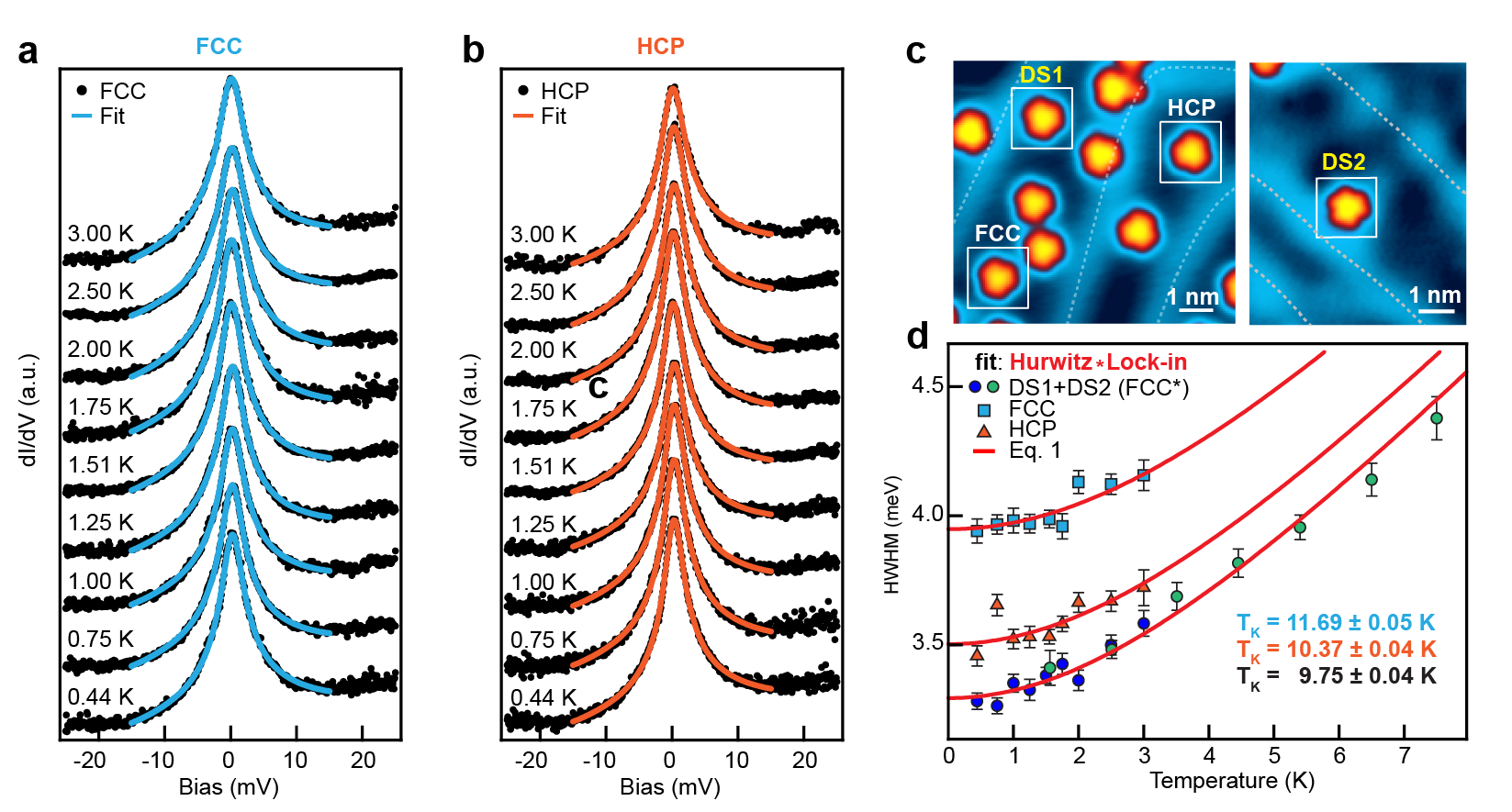}
  \caption{\label{fig:FittingHCPFCC}
    Experimental temperature-dependent $\didv$ spectra (black dots) of phenalenyl adsorbed on the \textit{fcc} (a) and \textit{hcp} (b) region of the Au(111) herringbone reconstruction with corresponding fits (solid lines) using lock-in-convoluted Hurwitz-Fano lineshape. Open feedback parameters: $V = -25$ mV, $I$ = 500 pA; Lock-in modulation: $V_{\rm m}$ = 0.5\,mV.
    (c) STM images of phenalenyl molecules on Au(111) with boxes highlighting the investigated molecules. Scanning parameters: -2\,V/100\,pA (left) and -0.1\,V/50\,pA (right).
    (d) Intrinsic Kondo linewidths as a function of temperature for the different adsorption sites, extracted by lock-in-convoluted Hurwitz fits (a,b and Fig.~3a of main text). Red solid lines display fits of $\Gamma(T)$ (Eq.~1 in main text) to the three adsorption sites.
    }
\end{figure}

\newpage
\section{Testing the validity of our approach for other Kondo systems}

We apply our methodology to four distinct spin-1/2 Kondo systems and their temperature-dependent measurements reported in Refs. ~\cite{mishra_observation_2021, mishra_topological_2020, Khajetoorians:NNano:2015, zheng_engineering_2020}. Refs. \cite{mishra_observation_2021,mishra_topological_2020, zheng_engineering_2020} feature all-carbon 
Kondo systems adsorbed on Au(111), while Ref. \cite{Khajetoorians:NNano:2015} discusses Fe adatoms on a Pt(111) surface.

\begin{figure}[h!]
  \includegraphics[width=15cm]{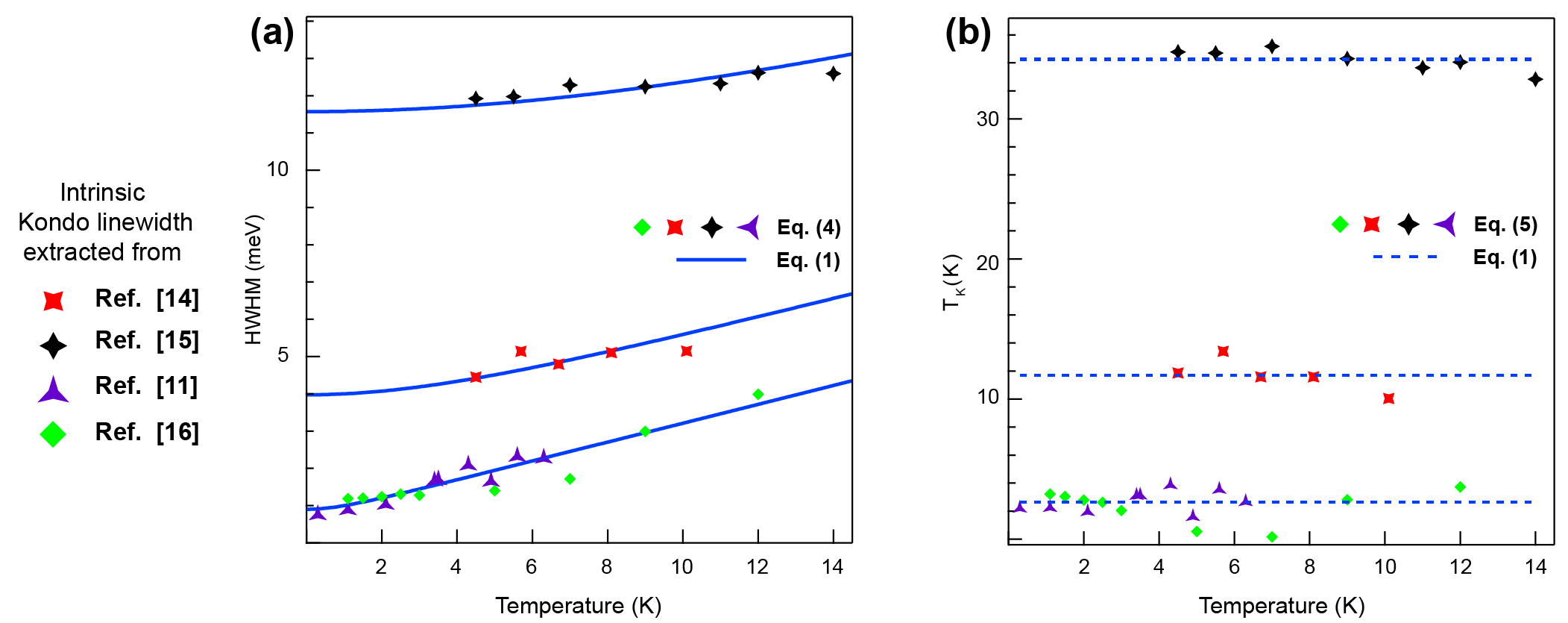}
  \caption{ 
    \label{Datasets}
    (a) \emph{Intrinsic} Kondo linewidths obtained from Hurwitz $\ast$ Lock-in fits versus temperature from datasets in Ref. \cite{mishra_observation_2021,mishra_topological_2020, Khajetoorians:NNano:2015, zheng_engineering_2020}.
    (b) For each \emph{intrinsic} Kondo linewidth, shown in (a), the corresponding $T_K$ is calculated via Eq. (5).  
}
\end{figure}

The experimental \textit{dI/dV} spectra were first fitted according to Eq. (4) (Hurwitz$\ast$Lock-in), and the obtained intrinsic HWHMs (shown in Fig. \ref{Datasets}a) were then fitted to Eq. (1). We also prove the validity of the \emph{shortcut} method (Fig. \ref{Datasets}b), where, for each intrinsic HWHM the corresponding $T_K$ is calculated according to Eq. (5) and compared to the value (dashed blue line) obtained from the fit to Eq. (1). Coincidentally the fitted Kondo temperatures of datasets in Ref. \cite{Khajetoorians:NNano:2015} and Ref. \cite{zheng_engineering_2020} are approximately equal. 

Figure \ref{Datasets}a convincingly demonstrates the validity of Eq.(1) for four distinct systems and temperature windows with respect to $\TK$. Panel b further justifies the use of the \emph{shortcut} method to rapidly evaluate $\TK$ from a single $dI/dV$ spectrum, and as a quick test to check if the system is properly thermalized during a temperature-dependent measurement.




\bibliography{refs,footnotes} 